\documentclass{PoS}
\usepackage{amsmath}
\usepackage{booktabs}
\usepackage{color}
\usepackage{subcaption}
\usepackage{bm}
\graphicspath{{figs/}}

\newcommand{\Vcb}{V_{cb}}
\newcommand{\Vub}{V_{ub}}
\newcommand{\Vcs}{V_{cs}}
\newcommand{\Vcd}{V_{cd}}
\newcommand{\bra}[1]{\langle #1 |}
\newcommand{\ket}[1]{| #1 \rangle}
\definecolor{dkgreen}{rgb}{0.0,0.4,0.0}

\def\[{\left[}
\def\]{\right]}
\def\({\left(}
\def\){\right)}
\newcommand{\matrixel}[3]{\left< #1 \vphantom{#2#3} \right|
 #2 \left| #3 \vphantom{#1#2} \right>} 
 
\DeclareMathOperator{\arccosh}{arcCosh}

\title{$B$- and $D$-meson semileptonic decays with highly improved staggered quarks}

\ShortTitle{$B$ and $D$ semileptonic decays with HISQ}

\author{\speaker{William I. Jay}$^{,1,2}$ and \speaker{Andrew Lytle}$^{,3}$\\
    E-mail: \email{willjay@mit.edu}\\
    E-mail: \email{atlytle@illinois.edu}\\
    }
    
\author{
Carleton DeTar$^4$,
Aida El-Khadra$^3$,
Elvira G{\'a}miz$^5$,
Zechariah Gelzer$^3$,
Steven Gottlieb$^6$,
Andreas Kronfeld$^2$,
Jim Simone $^2$,
and Alejandro Vaquero$^4$
\newline\newline
Fermilab Lattice and MILC Collaborations
\newline
\\
$^1$Center for Theoretical Physics, Massachusetts Institute of Technology, Cambridge, MA 02139, USA\\
$^2$ Theory Division, Fermi National Accelerator Laboratory, Batavia, Illinois, 60510, USA\\
$^3$ Department of Physics, University of Illinois at Urbana-Champaign, Urbana, Illinois, 61801, USA\\
$^4$ Department of Physics and Astronomy, University of Utah, Salt Lake City, Utah 84112, USA\\
$^5$ CAFPE and Departamento de Física Teórica y del Cosmos, Universidad de Granada, E-18071
Granada, Spain\\
$^6$ Department of Physics, Indiana University, Bloomington, Indiana 47405, USA\\
}

\abstract{
We present results for $B_{(s)}$- and $D_{(s)}$-meson semileptonic decays from ongoing calculations by the Fermilab Lattice and MILC Collaborations.
Our calculation employs the highly improved staggered quark (HISQ) action for both sea and valence quarks and includes several ensembles with physical-mass up, down, strange, and charm quarks and lattice spacings ranging from $a\approx0.15$ fm down to 0.06 fm.
At most lattice spacings, an ensemble with physical-mass light quarks is included.
The use of the highly improved action, combined with the MILC Collaboration's gauge ensembles with lattice spacings down to $a\approx0.042$ fm, allows heavy valence quarks to be treated with the same discretization as the light and strange quarks.
This unified treatment of the valence quarks allows (in some cases) for absolutely normalized currents, bypassing the need for perturbative matching, which has been a leading source of uncertainty in previous calculations of $B$-meson decay form factors by our collaboration.
All preliminary form-factor results are blinded.\\

MIT-CTP/5356
}

\FullConference{
    The 38th Annual International Symposium 
    on Lattice Field Theory - LATTICE2021\\
	26-30 July, 2021\\
	Zoom/Gather@MIT}

\begin{document}
\section{Introduction} \label{sec:Introduction}
Semileptonic decays of mesons are a rich source of information for 
understanding fundamental physics.
These relatively simple decay processes are especially useful for extracting CKM matrix elements,
since the decay rates are proportional to the square of the matrix element~\cite{HFLAV:2019otj,Donoghue:1992dd}.
In particular, the most precise exclusive extractions 
of $|\Vcb|$ and $|\Vub|$ come from measurements of $B \to D^{(*)} \ell \nu$ and $B \to \pi \ell \nu$.
In the case of $D$ decays, $D \to K \ell \nu$ gives a precise constraint on $|\Vcs|$ while $D \to \pi \ell \nu$ can be used to constrain $|\Vcd|$.
Central to these determinations are experimental measurements of the process and first-principles calculations of hadronic form factors from lattice QCD~\cite{FlavourLatticeAveragingGroup:2019iem}.

\begin{figure}
    \centering
    \includegraphics[width=0.49\textwidth]{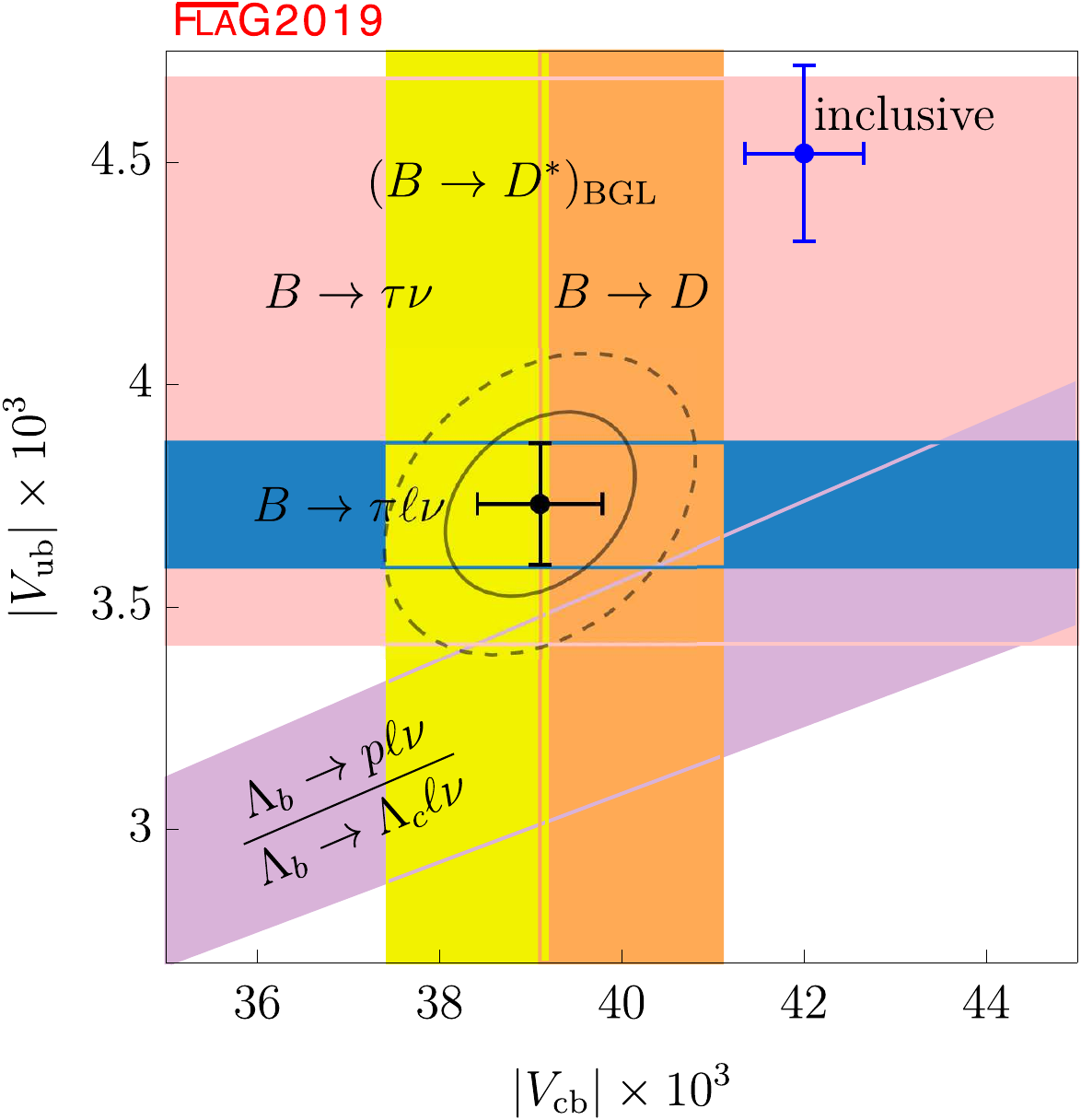}
    \includegraphics[width=0.49\textwidth]{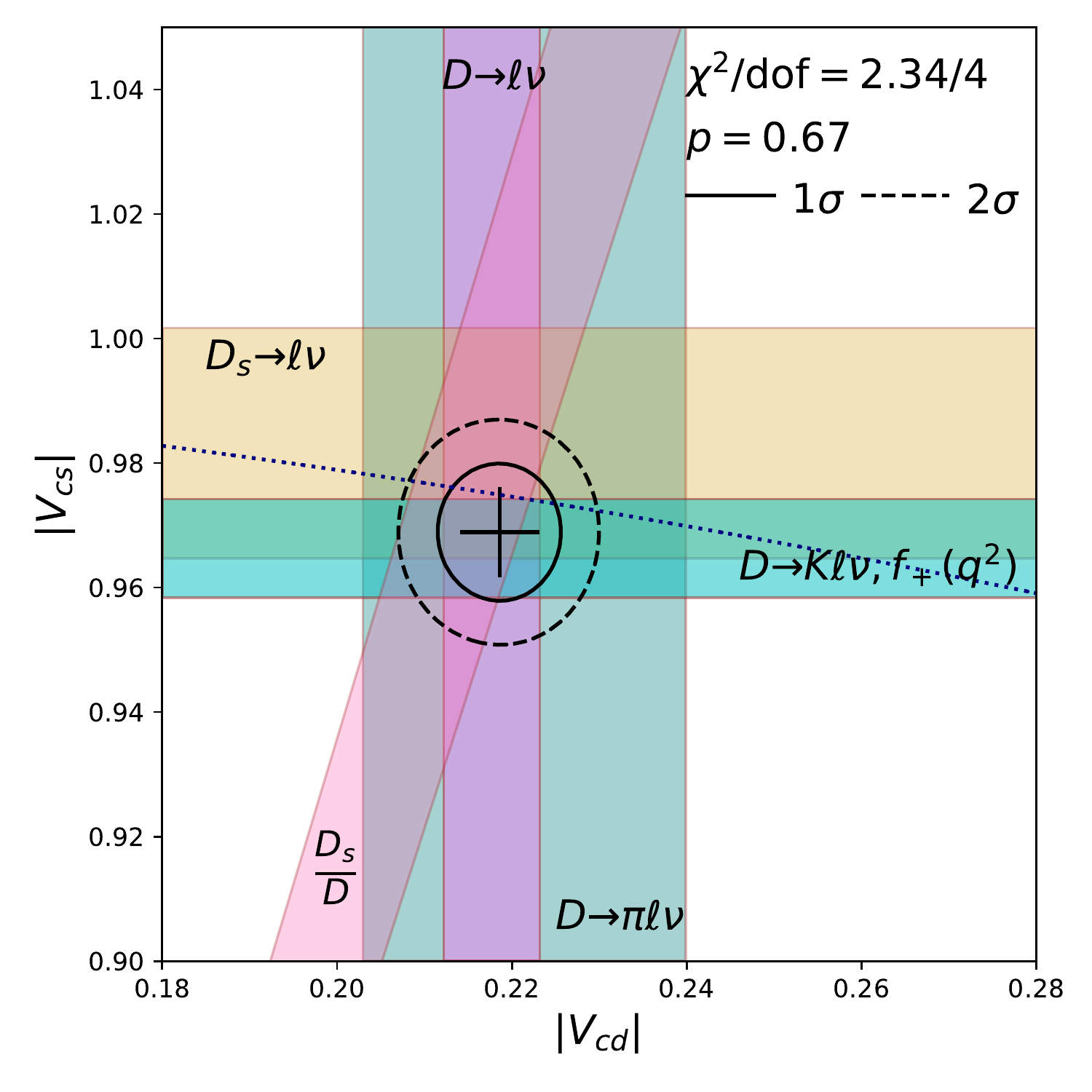}
    \caption{
    Status summaries of determinations of 
    $|\Vub|$, $|\Vcb|$, $|\Vcd|$, and $|\Vcs|$ from~\cite{FlavourLatticeAveragingGroup:2019iem} and~\cite{Wingate:2021ycr}.
    The colored bands give the CKM determination based on specific
    processes.
    The calculations described in this proceeding will lead to
    precision determinations for all four of these CKM matrix
    elements.}
    \label{fig:vub_vcb}
\end{figure}

In order to test the Standard Model as stringently as possible, one desires many independent determinations of a given CKM element.
Agreement among determinations signifies the validity of the Standard Model (for these processes, at this level of precision), whereas discrepancies could indicate the presence of new physics.
In the case of $|\Vcb|$ and $|\Vub|$, there are long-standing discrepancies between so-called inclusive determinations, which study decay rates summed over outgoing final states, and the most precise exclusive determinations from semileptonic decays~\cite{FlavourLatticeAveragingGroup:2019iem,Bouchard:2019all,Wingate:2021ycr}. 
This situation is summarized succinctly in Fig.~\ref{fig:vub_vcb}.

In addition to the outstanding inclusive/exclusive puzzles, there is by now a large body of evidence of so-called ``flavor anomalies'' uncovered at the $B$-factories and by LHCb (see, e.g.,~\cite{HFLAV:2019otj,BaBar:2019vpl,Belle:2018ezy,LHCb:2021lvy}).
Perhaps the most striking of these are in the $B \to D \ell \nu$ and $B \to D^* \ell \nu$ $R$-ratios, given simply as the ratio of branching fractions into the $\ell = \tau$ final state over $\ell = \mu, e$.
Lattice QCD calculations furnish all form factors needed to describe these decays and so can give predictions for integrated quantities like the $R$-ratios.
Therefore, lattice data are critical to anchor the expected distributions based on the Standard Model, and if these anomalies persist, to help understand and disentangle in detail the nature of new physics.

Here we adopt a unified approach to semileptonic decay calculations in which all quarks, both valence and sea, are treated using the same action. 
This setup allows us to compute the correlation functions needed for the whole gamut of pseudoscalar to pseudoscalar transitions simultaneously, resulting in a suite of form factors and improved determinations for all the aforementioned CKM elements.
Having a common setup for all the processes will allow us to study ratios of observables for different channels including all correlations; we expect that this capability will yield interesting results for comparison with experimental results.
As will be discussed in Sec.~\ref{ssec:lattice_calculation_details}, the combination of very fine lattice spacings with an improved staggered quark action is what makes this setup possible.

The remainder of this article is organized as follows.
In Sec.~\ref{sec:setup} we will briefly review the theory of semileptonic transitions with pseudoscalar initial and final states, and explain both the theory and relevant implementation details of our lattice QCD calculations.
In Sec~\ref{sec:fits} we present preliminary results.
Finally, we end with a few summary remarks in Sec~\ref{sec:summary}.

\clearpage

\section{Setup} \label{sec:setup}

\subsection{Theory Recap}

Matrix elements furnish a variety of form factors, defined according to
\begin{align}
\bra{L} \mathcal{V}^\mu \ket{H}
	&\equiv \sqrt{2 M_H} \left( v_H^\mu f_\parallel(q^2) + p_\perp^\mu f_\perp(q^2) \right)\\
	&\equiv f_+(q^2) \left( p_H^\mu + p_L^\mu - \frac{M_H^2 - M_L^2}{q^2} q^\mu \right) + f_0(q^2) \frac{M_H^2 - M_L^2}{q^2}q^\mu\\
\bra{L} \mathcal{S} \ket{H}
	&= \frac{M_L^2 - M_H^2}{m_h - m_\ell} f_0(q^2)	
\end{align}
In these expressions, $M_H$, $M_L$, $p_H^\mu$, and $p_L^\mu$ refer to the mass and four-momentum of the ``heavy" mother (H) and ``light" daughter (L) mesons; $m_h$ and $m_\ell$ refer to the ``heavy" mother and ``light" daughter quarks; $v_H^\mu = p_H^\mu / M_H$ is the four-velocity of the heavy meson; $p_\perp^\mu = p_L^\mu - (p_L \cdot v_H) v_H^\mu$ is the component of the daughter hadron's momentum orthogonal to $v_H$, and; $q^\mu = p_H^\mu - p_L^\mu$ is the momentum transfer.
The final equality relating $f_0$ to the scalar matrix element follows from partial conservation of the vector current (see Eq.~(\ref{eq:PCVC}) below).
The manifestly covariant expressions simplify in the rest frame of the decaying heavy meson and take the following forms:
\begin{align}
f_\parallel	&= Z_{V^0}\frac{\matrixel{L}{V^0}{H}}{\sqrt{2 M_H}} \label{eq:f_parallel}\\
f_\perp		&= Z_{V^i}\frac{\matrixel{L}{V^i}{H}}{\sqrt{2 M_H}} \frac{1}{p^i_L} \label{eq:f_perp}\\
f_0				&= Z_{S}\frac{m_h-m_\ell}{M_H^2 - M_L^2} \matrixel{L}{S}{H}. \label{eq:f_0}
\end{align}
These expressions are written in terms of bare lattice currents and their associated renormalization factors, so that $Z_\text{J} J_\text{lattice}$ has the same matrix elements (up to controlled uncertainties) as $\mathcal{J}$. 
The scalar and vector currents admit non-perturbative renormalization using partial conservation of the vector current (PCVC).
As an operator statement in the continuum, PCVC says $\partial_\mu V^\mu = (m_h - m_\ell)S$.
In momentum space, this result implies the following non-perturbative renormalization condition for the lattice currents~\cite{Na:2010uf}:
\begin{align}
    Z_{V^0}(M_H - E_L) \matrixel{L}{V^0}{H}
    + Z_{V^i} \bm{q}\cdot \matrixel{L}{\bm{V}}{H}
    = Z_{S} (m_h - m_\ell) \matrixel{L}{S}{H} \label{eq:PCVC}.
\end{align}
With the present treatment of all valence quarks with the highly improved staggered quark (HISQ) action~\cite{Follana:2006rc}, the local scalar density enjoys absolute normalization, $Z_S \equiv 1$.

Our ongoing calculations are also computing the tensor form factor, which is relevant in searches for physics beyond the Standard Model. 
However, in keeping with the scope of the talks actually presented at the conference, we restrict our discussion to the scalar and vector form factors.

\subsection{Lattice Calculation Details}\label{ssec:lattice_calculation_details}

\begin{table}[!htb]
\centering
\begin{tabular}{l l l }
$\approx a$ [fm] &   $m_\ell$    &   $m_h / m_c$ \\
\hline \hline
0.15   &   physical    & $0.9, 1.0, 1.1$ \\
0.12   &   physical    & $0.9, 1.0, 1.4$ \\
0.088   &   physical    & $0.9, 1.0, 1.5, 2.0, 2.5$ \\
0.088   &   $0.1 \times m_s$ & $0.9, 1.0, 1.5, 2.0, 2.5$ \\
0.057   &   physical    & $0.9, 1.0, 1.1, 2.2, 3.3$ \\
0.057   &   $0.1 \times m_s$ & $0.9, 1.0, 2.0, 3.0, 4.0$ \\
0.057   &   $0.2 \times m_s$ & $0.9, 1.0, 2.0, 3.0, 4.0$ \\
0.042   &   $0.2 \times m_s$ & $0.9, 1.0, 2.0, 3.0, 4.0, 4.2$
\end{tabular}
\caption{
	A summary of the lattice spacings and valence quark masses used in our calculation.
	The text describes the ensembles' sea and valence masses in more detail.
	The gauge ensembles were generated by the MILC collaboration
	~\cite{MILC:2010pul,MILC:2012znn,Bazavov:2017lyh}.
	\label{table:ensembles}
}
\end{table}

Our calculation uses ensembles generated by the MILC Collaboration using $N_f=2+1+1$ flavors of dynamical sea quarks with the HISQ action~\cite{MILC:2010pul,MILC:2012znn,Bazavov:2017lyh}.
Table~\ref{table:ensembles} summarizes the ensembles we have used in our calculation to date.
Lattice spacings range from $a\approx 0.15$~fm down to $a\approx 0.042$~fm.
An ensemble with physical-mass light quarks appears for all lattice spacings but $a\approx 0.042$ fm.
(For the decays of $B$ mesons, we plan to include a physical-mass ensemble at $a\approx 0.042$ fm in the future.) 
At the finest lattice spacings, we also include ensembles with heavier-than-physical light quarks with $m_\ell \approx 0.1 m_\text{strange}$ and $m_\ell \approx 0.2 m_\text{strange}$.
For the light and strange quarks, the valence quark masses match those in the sea.
In all ensembles the charm and strange quarks in the sea have physical masses.
The heavy valence quarks range in mass from roughly $0.9 m_c$ to just below the lattice cutoff $m_h a \lesssim 1$.
At the finest lattice spacing employed to date, $a\approx 0.042$~fm, this setup allows simulation close to the physical mass of the bottom quark.

\begin{figure}[t!]
    \centering
    \includegraphics[width=0.49\textwidth]{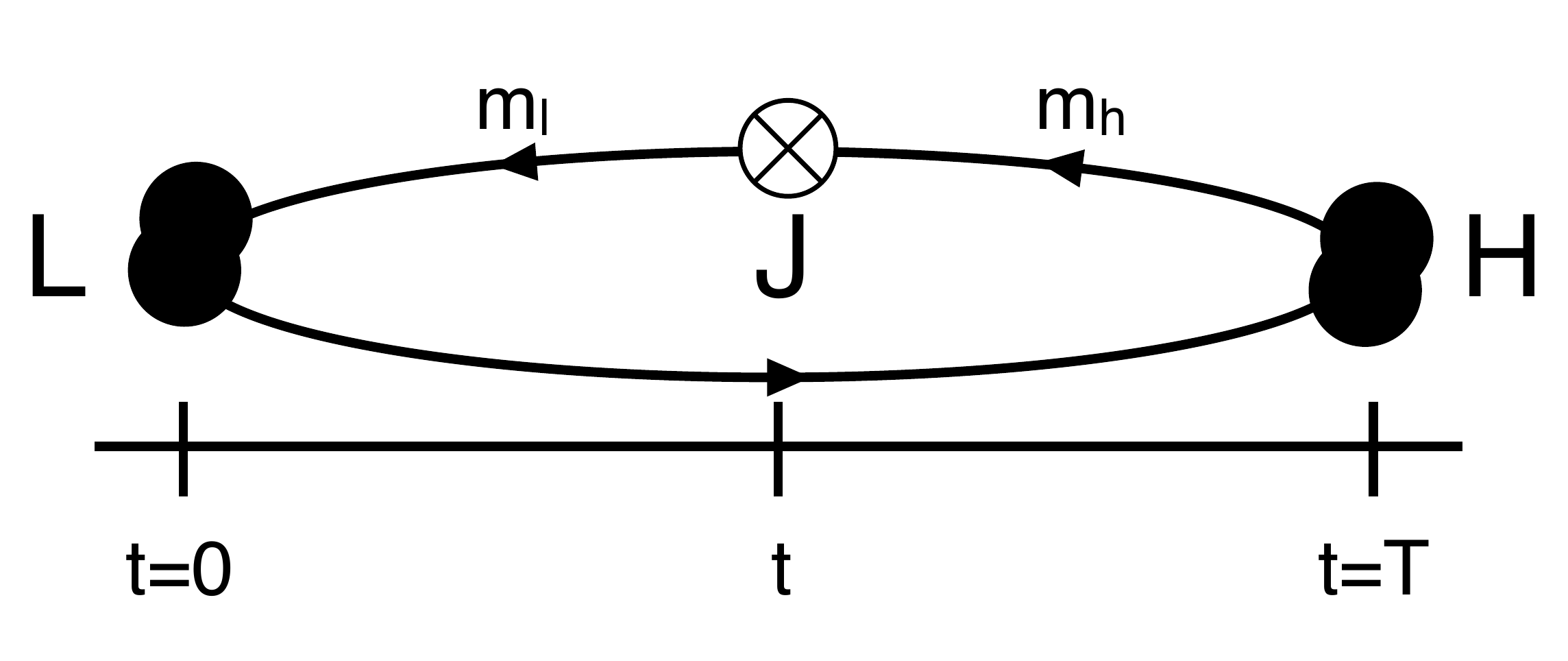}
    \caption{A schematic figure of the 3pt functions defined in Eq.~(\ref{eq:3pt}).
    The ``light" daughter hadron is created with momentum $\bm{p}$ at the origin.
    An external current $J$ is inserted at time $t$.
    The ``heavy" mother hadron is destroyed at rest at time $T$.
    }
    \label{fig:schematic_3pt}
\end{figure}

To access the form factors, we compute the following two-point and three-point correlation functions
\begin{align}
    C_H(t)
    &= \sum_{\bm{x}}
    \left\langle
        \mathcal{O}_H(0, \bm{0})
        \mathcal{O}_H(t, \bm{x})
    \right\rangle \label{eq:mother_2pt}\\
    C_L(t,\bm{p})
    &= \sum_{\bm{x}}
    e^{i \bm{p}\cdot\bm{x}}
    \left\langle
        \mathcal{O}_L(0, \bm{0})
        \mathcal{O}_L(t, \bm{x})
    \right\rangle \label{eq:daughter_2pt}\\
    C_3(t,T,\bm{p})
    &= \sum_{\bm{x},\bm{y}}
    e^{i \bm{p}\cdot\bm{y}}
    \left\langle
        \mathcal{O}_L(0, \bm{0})
        J(t, \bm{y})
        \mathcal{O}_H(T, \bm{x})
    \right\rangle, \label{eq:3pt}
\end{align}
where the $\mathcal{O}_{H,L}$ are staggered meson operators which couple to the mother $H$ and daughter $L$ hadrons.
For the scalar and temporal vector current we employ local staggered operators, while for spatial vector current we use the one-link operator.
For brevity, we have suppressed staggered structure and Lorentz indices in the lattice current $J$, which represents the scalar, vector, and tensor currents.
Figure~\ref{fig:schematic_3pt} shows the structure of these correlation functions.
We work in the rest frame of the decaying mother hadron $H$ and compute the recoiling daughter hadron $L$ with eight different lattice momenta $\bm{p}_L = \frac{2\pi}{N_s a} \bm{n}$, where $N_s \in \mathbb{Z}$ is the spatial extent of the lattice and $\bm{n}$ is $(0,0,0)$, $(1,0,0)$, $(1,1,0)$, $(2,0,0)$, $(2,1,0)$, $(3,0,0)$, $(2,2,2)$ or $(4,0,0)$.
For each choice of mass and momentum, we compute the three-point function for a few (typically 4 or 5) different source-sink separations $T$.
For the light-quark propagators in the calculation, we employ the truncated solver method~\cite{Bali:2009hu,Alexandrou:2012zz}, using 24 to 36 loose solves per configuration.

To extract the required matrix elements, our analysis employs joint correlated fits to the two-point and three-point correlation functions using the spectral decomposition.
For instance, for the three-point function in Eq.~(\ref{eq:3pt}), the spectral decomposition reads
\begin{align}
    C_3(t,T,\bm{p})
    = \sum_{m,n}
        (-1)^{m(t+1)} (-1)^{n(T-t-1)}
        A_{mn}
        e^{-E_L^{(n)}(\bm{p})t}
        e^{-M_L^{(m)}(T-t)}.\label{eq:3pt_spectral_decomp}
\end{align}
As usual for staggered fermions, the correlation functions include smoothly decaying contributions with the desired parity as well as oscillating contributions from states of opposite parity.
The spectral decompositions for the two-point functions are similar.
The ground-state amplitude $A_{00}$ is proportional to the matrix element $\matrixel{L}{J}{H}$, and so a fit to Eq.~(\ref{eq:3pt_spectral_decomp}) gives the required matrix element.
For the sake of visualization, the following ratio of correlation functions is useful:
\begin{align}
R^S(t, T, \bm{p}) =
    \sqrt{2 M_H} \left(\frac{m_h - m_l}{M_H^2 - M_L^2}\right)
    \frac{ C^S_3(t,T, \bm{p})}{ \sqrt{C_2^L(t, \bm{p}) C_2^H(T-t) e^{-E_L t} e^{-M_H (T-t)}}}.
\label{eq:ratio}
\end{align}
Up to discretization effects, this ratio asymptotically approaches the form factor $f_0$ for large times:
\begin{align}
    R^S(t, T, \bm{p}) \stackrel{0 \ll t \ll T}{\longrightarrow} f_0(\bm{p}).
\end{align}
Slightly different ratios, differing only by kinematic prefactors and renormalization factors, can also be constructed for $f_\parallel$ and $f_\perp$.
Although our quantitative analysis is based on fits to the spectral decomposition, the ratio provides a valuable visual check on the results.

The present analyses are blinded.
More precisely, all of our three-point correlation functions are multiplied by a blinding factor, $C^\text{Blind}_3(t,T,\bm{p}) = X \times C_3(t,T,\bm{p})$, where $X \in (0.95, 1.05)$ is a fixed random number.
Each sub-analysis ($D$ decays, $B$ decays) employs its own blinding factor.
We plan to carry the analysis of the blinded form factors all the way through the chiral interpolation and continuum extrapolation, unblinding only when the analysis of systematic errors is complete.
In these proceedings, all results involving three-point functions are blinded.

Several analysis choices are required when fitting the correlators.
The starting times $t_{\rm{min}}$ for the fits are taken to be a fixed physical distance across ensembles, $t_{\rm{min}} \approx 0.5$ fm.
Excluding data with noise-to-signal ratio greater than 30\% determines the value of $t_{\rm{max}}$ (for many correlators, this amounts to retaining all the data out to the midpoint of the lattice).
We fix the number of states included in each channel, typically 3 decaying and 1 oscillating (``$3+1$") states for the daughter meson and $3+2$ states for the mother meson.
The precise values are determined by analyzing the two-point functions in isolation and looking for the fewest states necessary to describe the correlator well for $t \gtrsim 0.5$ fm.
After fixing these parameters for our preferred analyses, we vary these choices and look at the variation of the matrix elements and energies.

\section{Results} \label{sec:fits}

\subsection{Two-point correlators}

\begin{figure}[t!]
    \centering
    \includegraphics[width=1.0\textwidth]{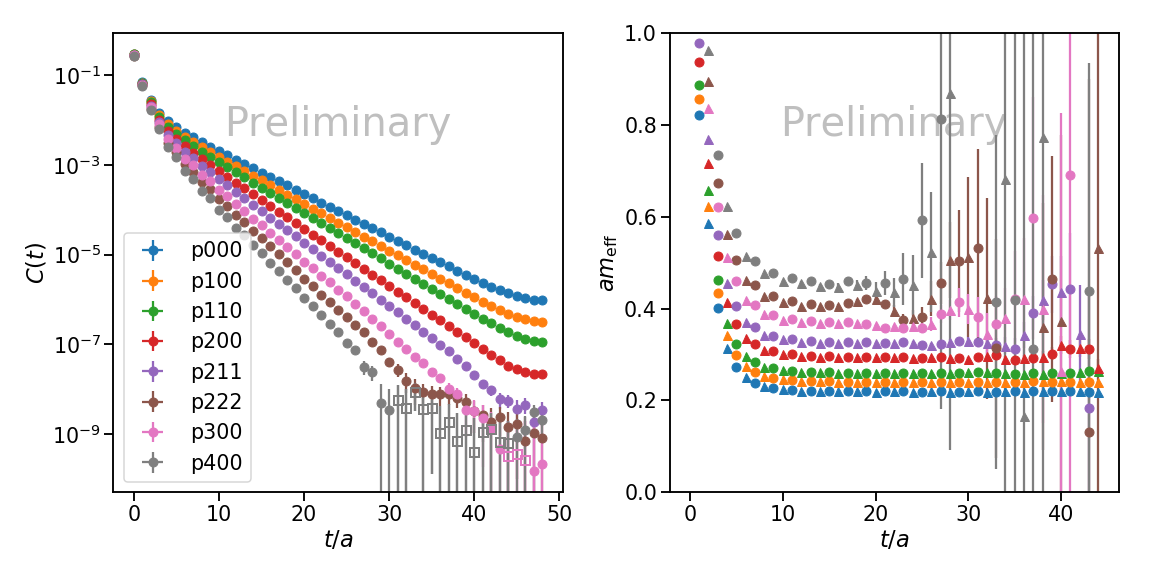}
    \caption{
        \textbf{Left:} Kaon two-point correlation functions on the physical-mass $a\approx0.09$ fm ensemble.
        \textbf{Right:} Effective masses.
        In both panels, the matched colors denote different momenta.
        }
    \label{fig:k_2pt_summary}
\end{figure}

\begin{figure}[t!]
    \centering
    \includegraphics[width=0.5\textwidth]{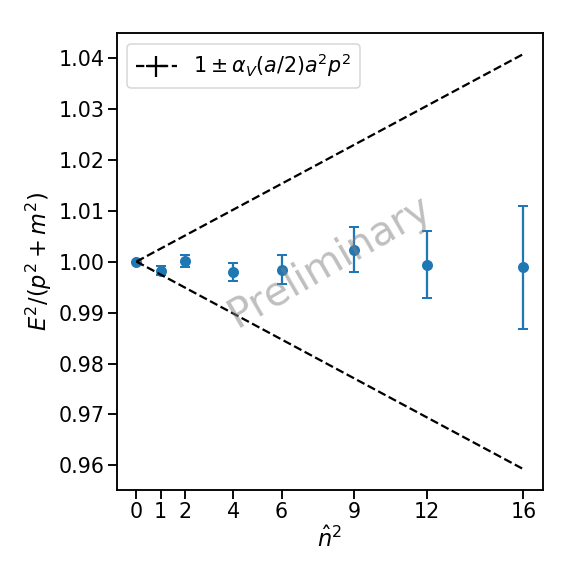}
    \caption{
        The ground-state energies from fits the correlators displayed in Fig~\ref{fig:k_2pt_summary} compared with continuum dispersion relation $E^2 = p^2 + m^2$.
        The envelope given by the black dotten lines denotes the order of magnitude of the expected discretization errors, $O(\alpha_s \, a^2 p^2)$.
        }
    \label{fig:k_dispersion}
\end{figure}

Figures~\ref{fig:k_2pt_summary} displays kaon two-point functions on the physical-mass $a\approx 0.09$ fm ensemble.
These correlators are fairly representative of the quality of two-point functions across masses and lattice spacings.
As expected, the noise-to-signal ratio grows exponentially for boosted correlators, and so highly boosted correlators, e.g. with $\bm{n}=(2,2,2)$ or $(4,0,0)$, become noisy at large times.
Even so, clear plateaus spanning several time slices are present in the effective mass curves at each momentum in the center pane.\footnote{
To reduce the visual impact of the opposite parity states, we construct the effective masses separately on even and odd time slices using $am_{\rm eff} \equiv \frac{1}{2}\arccosh \left[ (C(t+2)+ C(t-2))/C(t) \right]$.
The triangle and circle markers in the middle pane of Fig.~\ref{fig:k_2pt_summary} correspond to the even and odd time slices, respectively.
}
Fig.~\ref{fig:k_dispersion} shows the ground-state energies from fitting the correlators in Fig~\ref{fig:k_2pt_summary}.
The good agreement with the continuum dispersion relation, $E^2 = \bm{p}^2 + m^2$, demonstrates the good control of discretization effects afforded by the use the HISQ action.  
Using the results of these and other fits to two-point functions, we fix $t_{\rm{min}} \approx 0.5$ fm and the number of states for subsequent fits involving three-point functions.

\subsection{$D$-meson decays}

\begin{figure}[t!]
    \centering
    \includegraphics[width=1.0\textwidth]{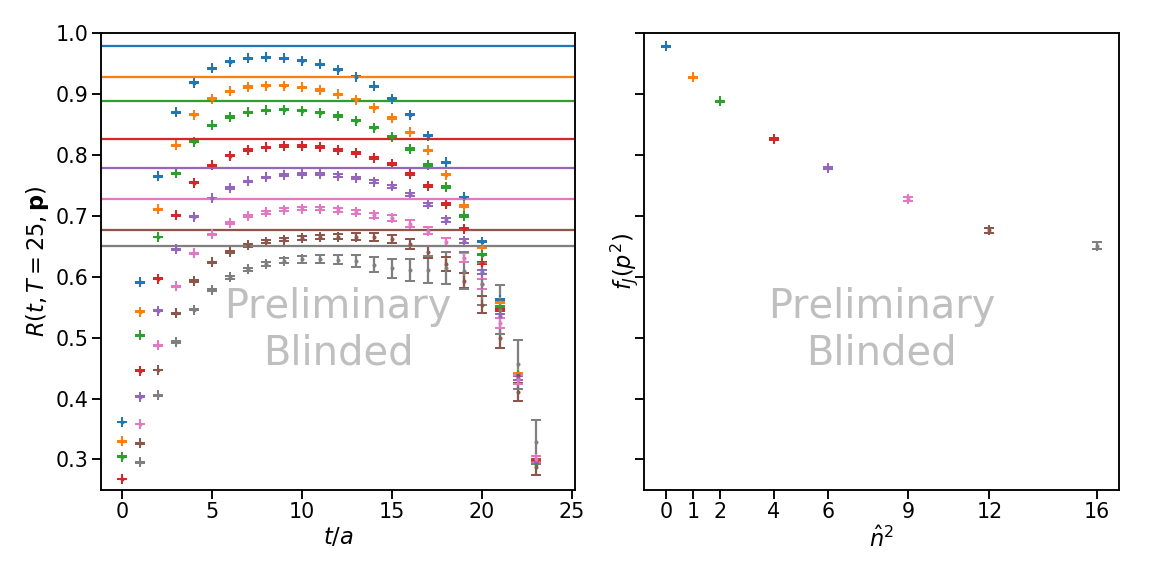}
    \caption{
    \textbf{Left:} Best-fit results (solid horizontal bands) for the bare form factor $f_0^{Ds \to K}$ on the physical-mass $a\approx 0.09$ fm ensemble.
    The points show the ratio Eq.~(\ref{eq:ratio}) for fixed source-sink separation $T=25$.
    The colors correspond to different momenta and match the conventions of Fig.~\ref{fig:k_2pt_summary}.
    \textbf{Right:} The momentum dependence of $f_0^{Ds \to K}$ on the same ensemble.
    }
    \label{fig:Ds2K_summary}
\end{figure}

Figure~\ref{fig:Ds2K_summary} summarizes fit results for the scalar form factor $f_0$ for the transition $D_s \to K$ on the physical-mass $a\approx 0.09$~fm ensemble.
The left pane shows data for the ratio defined in Eq.~(\ref{eq:ratio}) together with the horizontal bands showing best-fit values for the form factors at each momentum.
The right pane shows the behavior of the form factor as a function of momentum.
The general results are qualitatively similar for the other lattice spacings, currents, and transitions.

\begin{figure}[t!]
    \centering
    \includegraphics[width=0.49\textwidth]{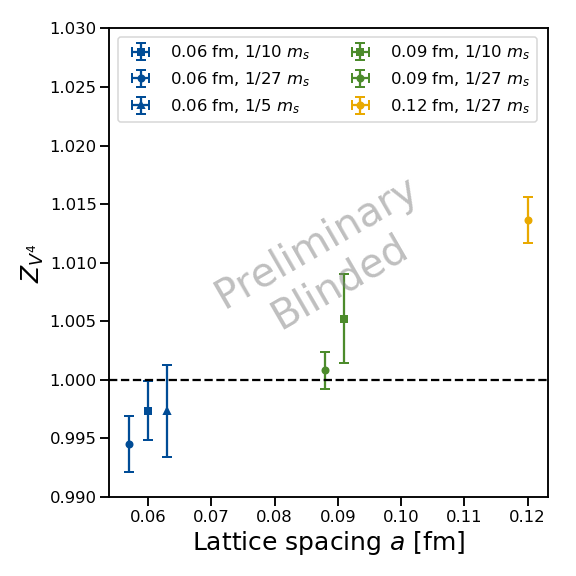}
    \includegraphics[width=0.49\textwidth]{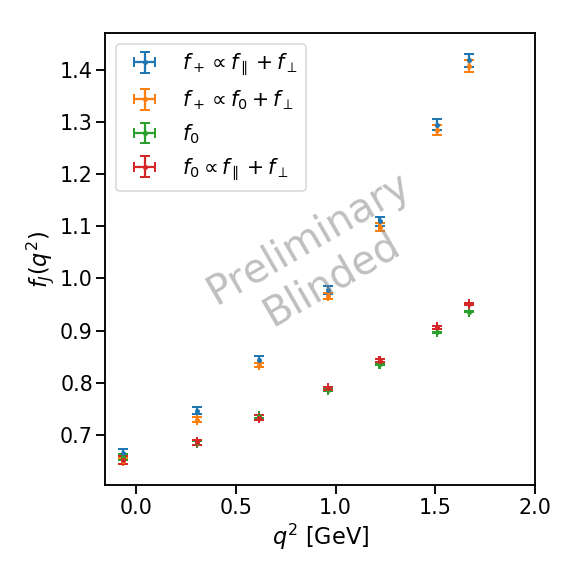}
    \caption{
    \textbf{Left:} Results for the renormalization factor $Z_{V^4}$, defined in Eq.~(\ref{eq:PCVC}), across several ensembles.
    \textbf{Right:} Results for the renormalized form factors $f_+$ and $f_0$ constructed in different ways from $f_0$, $f_\parallel$ and $f_\perp$.
    }
    \label{fig:renormalization}
\end{figure}

Renormalization of the vector current is done non-perturbatively using PCVC.
In principle, much freedom exists for extracting the Z-factors using Eq.~(\ref{eq:PCVC}).
The present analysis fits the bare matrix elements as a function of momentum to Eq.~(\ref{eq:PCVC}), treating $Z_{V^4}$ and $Z_{V^i}$ as fit parameters.
The left pane of Fig.~\ref{fig:renormalization} shows the results of this procedure for the vector-current renormalization factor $Z_{V^4}$ across the different ensembles; the results for $Z_{V^i}$ are similar.
Because three-point functions of $V^4$, $\bm{V}$, and $S$ are all multiplied by the same blinding factor, the fitted $Z_{V^4}$ and $Z_{V^i}$ are not affected by the blinding procedure.

Our analysis directly computes the renormalized form factors $f_0$, $f_\parallel$, and $f_\perp$ from the relations in Eqs.~(\ref{eq:f_parallel})--(\ref{eq:f_0}).
Most relevant for phenomenology are the form factors $f_0$ and $f_+$, the latter of which can be obtained from a linear combination of $f_\parallel$ and $f_\perp$.
Besides the relation involving the scalar matrix element in Eq.~(\ref{eq:f_0}), the scalar form factor $f_0$ can also be reconstructed from $f_\parallel$ and $f_\perp$.
Moreover, a kinematic constraint says that $f_+=f_0$ at the point of vanishing momentum transfer, $q^2=0$.
Together, these relations offer the chance to perform consistency checks by constructing $f_0$ and $f_+$ in different ways.
The right pane of Figure~\ref{fig:renormalization} shows the result of such checks for the transition $D_s\to K$ on the physical-mass $a\approx 0.09$~fm ensemble, with excellent agreement for both constructions of $f_+$ (upper set of points) and $f_0$ (lower set of points).
As expected, all agree near $q^2=0$.

\subsection{$B$-meson decays}
Although the physical-mass charm quark can be comfortably accommodated on all of our lattice ensembles (even at relatively coarse lattice spacings), the lattice cutoff $1/a$ is below the physical mass of the $b$ quark until one reaches our (current) finest lattice spacing of $a \approx 0.042$ fm, for which $am_b \approx 0.84$.
In order to address this issue, on each ensemble we work at a range
of heavy quark mass values between charm and bottom, limited to
$am_h \lesssim 1$. 
These values are listed in Table~\ref{table:ensembles} in units of $m_c$.
In this way we can map out the
dependence of our form factors as the quark mass varies between
charm and bottom.

\begin{figure}
    \centering
    \includegraphics[width=\textwidth]{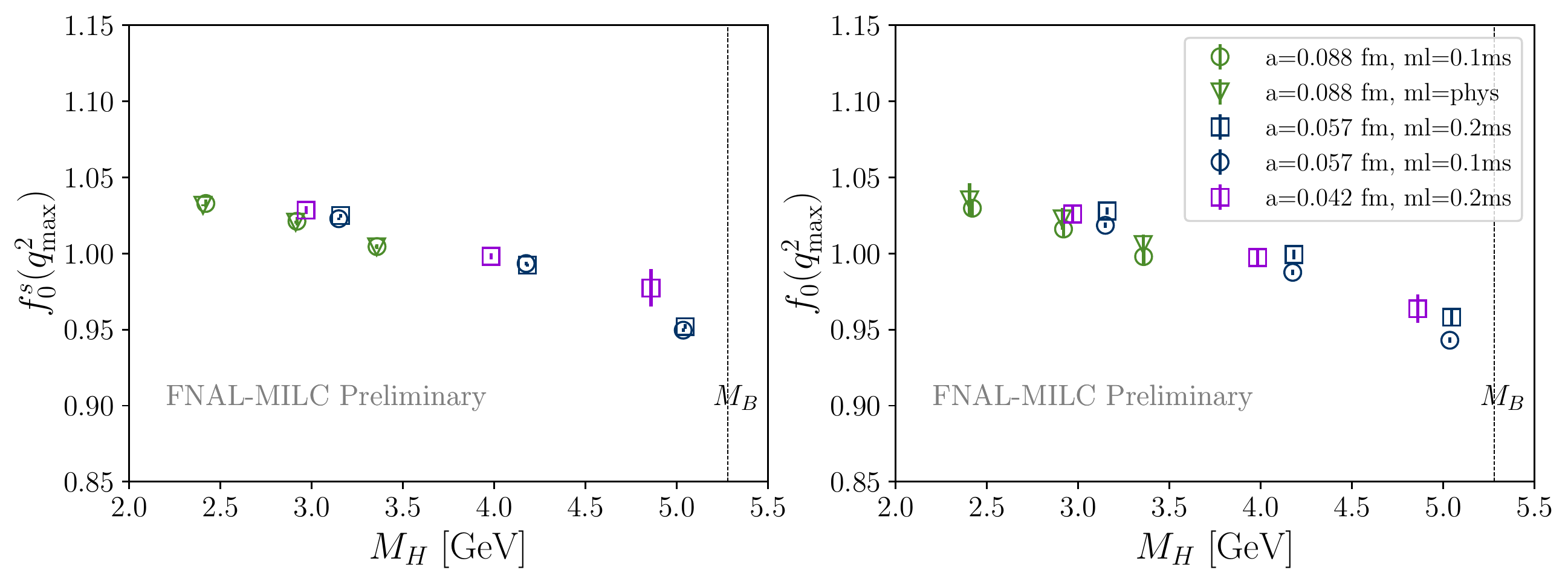}
    \caption{
    $B_s \to D_s$ (left) and $B \to D$ (right) zero recoil form factor 
    results. The data have been improved by removing 
    dominant tree-level
    discretization artifacts, so that the resultant data 
    roughly trace out
    a curve corresponding to the physical variation of $f_0(q^2_{\text{max}})$ with heavy quark input mass $m_h$.
    Note that for $B_s \to D_s$ the data points at the same lattice
    spacing but with different light sea-quark masses overlap.}
    \label{fig:f0_q2max_vs_MH}
\end{figure}

Examples of this procedure are shown in Fig.~\ref{fig:f0_q2max_vs_MH} for the $B_{(s)} \to D_{(s)}$ form factor $f_0$ at zero-recoil.
The $x$-axis in these figures is the mass $M_H$ of an ``$H$'' meson, a physical proxy for the heavy quark, composed of an $h$-quark and a light quark, with the target $B$ mass on the right of the figure. 
One can see that as we go to finer lattices we are able access masses closer to the physical $B$ meson.
The raw data coming from correlator fits has been improved in these figures by removing known tree-level discretization artifacts~\cite{Bazavov:2017lyh} in the extraction of $f_0$.
Evidently, these factors represent the bulk of the heavy mass discretization effects, and so after their removal the data have largely collapsed onto a curve representing the actual evolution of $f_0$ as heavy quark mass is varied.
Additional variation stems from a combination of discretization artifacts, light sea-quark, and mistuning effects that will be the subject of future analysis.

\begin{figure}
    \centering
    \includegraphics[width=\textwidth]{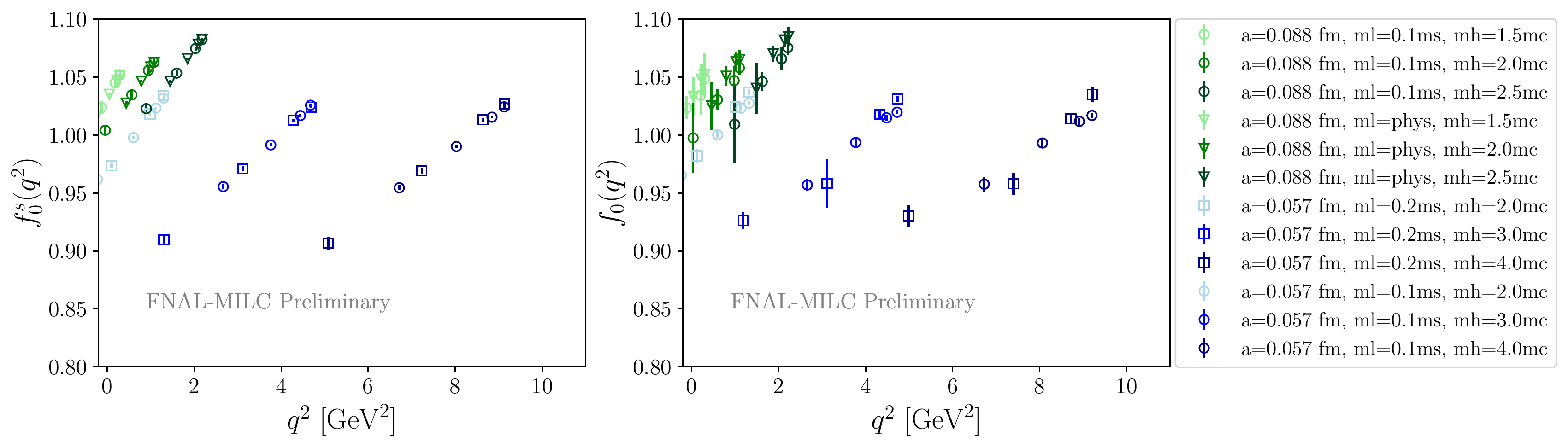}
    \caption{$B_s \to D_s$ (left) and $B \to D$ (right) $f_0(q^2)$ form factor results at zero and non-zero recoil
    momentum. Note the small statistical errors, 
    particularly for the $B_s \to D_s$ case with no
    valence light quarks, and good coverage over the $q^2$ range
    of the decays.}
    \label{fig:f0_vs_q2}
\end{figure}

Figure~\ref{fig:f0_vs_q2} shows preliminary results for the $f_0$
form factors now extended to non-zero recoil momentum 
($q^2 < q^2_{\text{max}})$ on the $a\approx0.09$ and 0.06 fm ensembles. 
In order to convey visually the size of the discretization effects in the raw data, the data in these figures have not been improved with tree-level discretization factors.
(Our final analysis will, of course, incorprorate these effects.)
Note that good statistical control is achieved even for large momentum values, and that good coverage is present over the physical $q^2$ range.

\section{Summary} \label{sec:summary}
We have presented preliminary results for $B_{(s)}$- and $D_{(s)}$-semileptonic decay form factors computed using the HISQ valence action on the HISQ $N_f=2+1+1$ ensembles generated by the MILC Collaboration.
These quantities are critical inputs in the quest to elucidate the flavor sector of the Standard Model and understand or resolve the nature of so-called flavour anomalies.
In particular, the calculations outlined here together with high-precision experimental measurements will result in improved determinations of $|\Vcs|$, $|\Vcb|$, $|\Vcd|$, $|\Vub|$, providing some of the most stringent tests of the Standard Model in these respective sectors.

Although our results are still preliminary, we observe good statistical control over the kinematic range studied.
Our calculation benefits from fine lattice spacings, which are especially critical for handling the $b$ quark, and ensembles with physical light quarks at most lattice spacings. 
This treatment means that our final chiral analysis will be an interpolation (instead of an extrapolation), which we expect to reduce systematic errors.
In the future, we will extend our calculation to ensembles with an even finer lattice spacing of $a\approx0.03$ fm.

\section*{Acknowledgments}
We are happy to acknowledge support and feedback from our friends and colleagues in the Fermilab Lattice and MILC Collaborations.

This document was prepared by the Fermilab Lattice and MILC Collaborations using the resources of the Fermi National Accelerator Laboratory (Fermilab), a U.S. Department of Energy, Office of Science, HEP User Facility.
Fermilab is managed by Fermi Research Alliance, LLC (FRA), acting under Contract No. DE-AC02-07CH11359.
This material is based upon work supported
by the U.S. Department of Energy, Office of Science, Office of Nuclear Physics under grant Contract Numbers DE-SC0011090 (W.J.), DE-SC0021006 (W.J.), DE-SC0015655 (A.L., A.X.K.), and DE-SC0010120 (S.G.); 
by the U.S. National Science Foundation under Grants No. PHY17-19626 and PHY20-13064 (C.D., A.V.);
by SRA (Spain) under Grant No.\ PID2019-106087GB-C21 / 10.13039/501100011033 (E.G.);
by the Junta de Andalucía (Spain) under Grants No.\ FQM- 101, A-FQM-467-UGR18, and P18-FR-4314 (FEDER) (E.G.).

Computations for this work were carried out in part on facilities of the USQCD Collaboration, which are funded by the Office of Science of the U.S. Department of Energy.
An award of computer time was provided by the Innovative and Novel Computational Impact on Theory and Experiment (INCITE) program. This research used resources of the Argonne Leadership Computing Facility, which is a DOE Office of Science User Facility supported under contract DE-AC02-06CH11357. This research also used resources of the Oak Ridge Leadership Computing Facility, which is a DOE Office of Science User Facility supported under Contract DE-AC05-00OR22725.
This research used resources of the National Energy Research Scientific Computing Center (NERSC), a U.S. Department of Energy Office of Science User Facility located at Lawrence Berkeley National Laboratory, operated under Contract No. DE-AC02-05CH11231.
The authors acknowledge support from the ASCR Leadership Computing Challenge (ALCC) in the form of time on the computers Summit and Theta.
The authors acknowledge the \href{http://www.tacc.utexas.edu}{Texas Advanced Computing Center (TACC)} at The University of Texas at Austin for providing HPC resources that have contributed to the research results reported within this paper.
This research is part of the Frontera computing project at the Texas Advanced Computing Center. Frontera is made possible by National Science Foundation award OAC-1818253~\cite{Frontera}.
This work used the Extreme Science and Engineering Discovery Environment (XSEDE), which is supported by National Science Foundation grant number ACI-1548562.
This work used XSEDE Ranch through the allocation TG-MCA93S002~\cite{XSEDE}.


\end{document}